# Calibration of chemical sensors in mobile wireless networks

**Rakesh Gosangi, Harsha Chenji, Radu Stoleru, and Ricardo Gutierrez-Osuna**
*Texas A&M University, College Station, Texas, USA*
rakesh, cjh, stoleru, rgutier@cse.tamu.edu

## ABSTRACT
Low-power chemical sensors deployed on mobile platforms make it possible to monitor pollutant concentrations across large urban areas. However, chemical sensors are prone to drift (e.g., aging, damage, poisoning) and have to be calibrated periodically. In this paper, we present an opportunistic calibration approach that relies on encounters between sensors; when in vicinity of each other, sensors exchange measurements and use the accumulated information to re-calibrate. We formulate the calibration process as weighted least-squares, where the most recent measurements are assigned the highest weights. We model the weights with an exponential decay function (in time) and optimize the decay constant using simulated annealing (SA). We validated the proposed method on a simulated sensor network with the sensors' mobility driven by random-waypoint (RWP) models. We present results in terms of average calibration errors for different weight functions, and network sizes.

## CALIBRATION OF MOBILE CHEMICAL SENSORS

Chemical sensors can be calibrated manually by exposing them to known stimuli. While this type of calibration works well with fixed instruments [1] it becomes impractical with mobile platforms. As a result, several collaborative calibration methods have been proposed [2-4] where sensors exchange measurements when in proximity; this allows sensors to recalibrate periodically, once a sufficient number of measurements have been exchanged. In this paper, we present a strategy to optimally weigh these measurements such that the more recent measurements are given higher importance in calibration process.

## METHODS

Consider a network of $n$ mobile gas sensors $\{s_1, s_2, \ldots, s_n\}$ continuously monitoring the concentration of a known chemical in the environment. Assume sensors are enabled with short-range wireless communication (e.g. Bluetooth), and their mobility driven by an RWP model [5]. Each sensor $s_i$ is prone to drift such that its raw output $y_i(t)$ is a nonlinear function of the ground truth $x(t)$:

$$y_i(t) = f_i\big(x(t), \gamma_i(t)\big) = \sum_{j=0}^{r} d_{i,j}(t) x(t)^j \qquad (1)$$

where $\gamma_i(t) = \big(d_{i,0}(t), d_{i,1}(t), \ldots, d_{i,r}(t)\big)$ is a set of $r$ drift parameters for sensor $s_i$. We assume the drift parameters themselves change linearly over time as: $d_{i,j}(t) = d_{i,j}(0)(1 + \alpha_{i,j} t)$. Thus, calibration entails estimating the drift parameters for all sensors. To solve this problem, our approach maintains a calibration table per sensor $CT_i = \big\{\big(y_i(t_1), \hat{x}(t_1)\big), \ldots, \big(y_i(t_p), \hat{x}(t_p)\big)\big\}$, where $y_i(t_k)$ is the raw sensor output and $\hat{x}(t_k)$ is the estimated ground truth. Whenever $s_i$ is in the vicinity of another sensor, a new tuple is added to $CT_i$ and the sensor is recalibrated. To estimate $\hat{\gamma}_i(t)$, we minimize the weighted sum of squared differences between calibrated values and raw sensor readings:

$$\gamma_i(t) = \arg\min_{\gamma_i(t)} \sum_{k=1}^{p} \omega_k \big(f_i\big(\hat{x}(t_k), \gamma_i(t_k)\big) - y_i(t_k)\big)^2 \qquad (2)$$

where $\hat{x}(t_k)$ is the weighted average of measurements from neighboring sensors: given sensors $s_1, s_2, \ldots, s_m$ in proximity at $t_k$, then $\hat{x}(t_k) = \sum_{j=1}^{m} \rho_j \hat{x}_j(t_k)$, where $\hat{x}_j(t_k)$ is the calibrated response of $s_j$, and $\rho_j$ is a weight that decays exponentially since the last calibration time $\tau_j$ for the sensor: $\rho_j \propto 1/(t_k - \tau_j)$.

Each tuple in eq. (2) is weighted according to its age $\omega_k = \exp(-\lambda_i(t - t_k))$; this ensures that the most recent tuples are given more importance in the re-calibration process. We optimize the decay constant $\lambda_i$ offline, before the sensor is deployed. First, we expose the sensor to a known set of stimuli (over a period of time) and collect the responses. Then, we use eq. (2) to estimate $\hat{\gamma}_i(t)$ for different values of $\lambda_i$. Finally, we use SA to find $\lambda_i$ with the lowest error.

## EXPERIMENTAL RESULTS

We tested the method on a simulated network with 30 mobile sensors (transmission range of 20m) distributed over a 1km$^2$ grid. We simulated the gas concentration as a 2-D Gaussian distribution across the grid (std. dev. $\sigma_x = \sigma_y = 50$m), with the position of its center driven by a random walk. The sensors were subjected to quadratic drift, with drift parameters initialized using uniform distributions $d_{i,0} \in (0,1.5)$, $d_{i,1} \in (2,3)$, and $d_{i,2} \in (1,2)$, and $\alpha_{i,j} = 0.002$ for all sensors. We ran simulations (10 repetitions, 1000 time steps each) with three different weight functions: (1) uniform: all tuples were weighted equally (i.e., $\lambda_i = 0$), (2) linear: weights were inversely proportional to the tuples' ages: $\omega_k \propto 1/(t - t_k)$, and (3) optimized: as in eq. (2). To ensure a fair comparison between these functions, we used the same traces for all the sensors. Results are shown in Fig. 1(a). As expected, the errors are highest when the sensors are uncalibrated. Uniform weights provide a relatively small (~3%) but not statistically significant improvement ($p=0.96$). Linear and optimized weights reduce the errors by 18% and 36% respectively, both significant results ($p<0.001$). We also characterized the performance of the method as a function of networks size (ranging from 5 to 50). Results are shown in Fig. 1(b). As the network size increases, so do the average number of rendezvous among the sensors. As a result, the sensors calibrate more frequently and the average errors reduce.

These results illustrate the importance of using appropriate weight functions for collaborative calibration. Our future work includes comprehensive evaluation with realistic mobility patterns and experimental data.

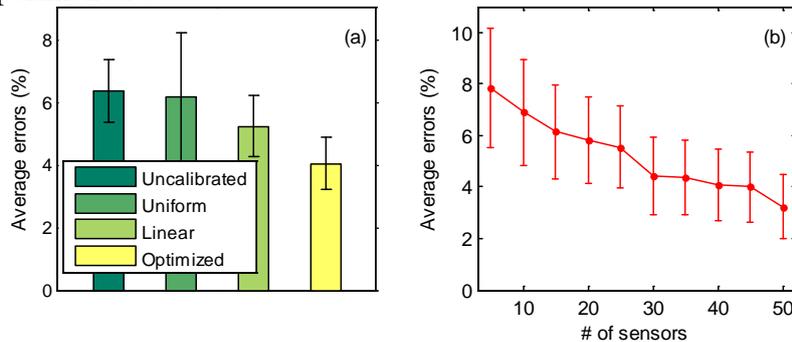

**Fig. 1. The average % error (across all sensors) for different (a) weight functions, and (b) network sizes.**